\documentclass[onecolumn,showpacs,amsmath,amssymb,12pt]{revtex4}
\usepackage{graphicx}
\usepackage{dcolumn}
\usepackage{bm}
\begin{document}
%\preprint{APS/123-QED}
%%%%%%%%%%%%%%%%%%%%%%%%
\newcommand{\hs}{\hspace*{0.5cm}}
\newcommand{\vs}{\vspace*{0.5cm}}
\newcommand{\be}{\begin{equation}}
\newcommand{\ee}{\end{equation}}
\newcommand{\bea}{\begin{eqnarray}}
\newcommand{\eea}{\end{eqnarray}}
\newcommand{\ben}{\begin{enumerate}}
\newcommand{\een}{\end{enumerate}}
\newcommand{\bde}{\begin{widetext}}
\newcommand{\ede}{\end{widetext}}
\newcommand{\nn}{\nonumber}
\newcommand{\crn}{\nonumber \\}
\newcommand{\Tr}{\mathrm{Tr}}
\newcommand{\non}{\nonumber}
\newcommand{\noi}{\noindent}
\newcommand{\al}{\alpha}
\newcommand{\la}{\lambda}
\newcommand{\bet}{\beta}
\newcommand{\ga}{\gamma}
\newcommand{\va}{\varphi}
\newcommand{\om}{\omega}
\newcommand{\pa}{\partial}
\newcommand{\+}{\dagger}
\newcommand{\fr}{\frac}
\newcommand{\bc}{\begin{center}}
\newcommand{\ec}{\end{center}}
\newcommand{\Ga}{\Gamma}
\newcommand{\de}{\delta}
\newcommand{\De}{\Delta}
\newcommand{\ep}{\epsilon}
\newcommand{\varep}{\varepsilon}
\newcommand{\ka}{\kappa}
\newcommand{\La}{\Lambda}
\newcommand{\si}{\sigma}
\newcommand{\Si}{\Sigma}
\newcommand{\ta}{\tau}
\newcommand{\up}{\upsilon}
\newcommand{\Up}{\Upsilon}
\newcommand{\ze}{\zeta}
\newcommand{\ps}{\psi}
\newcommand{\Ps}{\Psi}
\newcommand{\ph}{\phi}
\newcommand{\vph}{\varphi}
\newcommand{\Ph}{\Phi}
\newcommand{\Om}{\Omega}
%%%%%%%%%%%%%%%%%%%%%%%%

\title{Simple 3-3-1 model and implication for dark matter}

\author{P. V. Dong}
\email {pvdong@iop.vast.ac.vn}\affiliation{Institute of Physics, Vietnam Academy of Science and Technology,
 10 Dao Tan, Ba Dinh, Hanoi, Vietnam}
 
\author{N. T. K. Ngan}
\email{ntkngan@ctu.edu.vn}\affiliation{Department of Physics, Cantho University, 3/2 Street, Ninh Kieu, Cantho, Vietnam}

\author{D. V. Soa}
\email{dvsoa@assoc.iop.vast.ac.vn}\affiliation{Department of Physics, Hanoi National University of Education,
 136 Xuan Thuy, Cau Giay, Hanoi, Vietnam}

\date{\today}

\begin{abstract}

We propose a new and realistic 3-3-1 model with the minimal lepton and scalar contents, named the simple 3-3-1 model. The scalar sector contains two new heavy Higgs bosons, one neutral $H$ and another singly-charged $H^\pm$, besides the standard model Higgs boson. There is a mixing between the $Z$ boson and the new neutral gauge boson ($Z'$). The $\rho$ parameter constrains the 3-3-1 breaking scale ($w$) to be $w>460$ GeV. The quarks get consistent masses via five-dimensional effective interactions while the leptons via interactions up to six dimensions. Particularly, the neutrino small masses are generated as a consequence of the approximate lepton-number symmetry of the model. The proton is stabilized due to the lepton-parity conservation $(-1)^L$. The hadronic FCNCs are calculated that give a bound $w>3.6$ TeV and yield that the third quark generation is different from the first two. The correct mass generation for top quark implies that the minimal scalar sector as proposed is unique. By the simple 3-3-1 model, the other scalars beside the minimal ones can behave as inert fields responsible for dark matter. A triplet, doublet and singlet dark matter are respectively recognized. Our proposals provide the solutions for the long-standing dark matter issue in the minimal 3-3-1 model.

\end{abstract}


\maketitle

\section{\label{intro}Introduction}

The standard model has been extremely successful in describing observed phenomena, especially for the outstanding prediction of recently-discovered Higgs boson \cite{atlascms}. However, it must be extended to address unsolved questions such as small masses and mixing of neutrinos, matter-antimatter asymmetry of the universe, dark matter and dark energy \cite{pdg}. Therefore, we would like to argue that the $SU(3)_C\otimes SU(3)_L\otimes U(1)_X$ (3-3-1) gauge theory where the color group is as usual while the electroweak group is enlarged \cite{331m,331r,ecn331,r331} may be an interesting choice for the physics beyond the standard model, specially for the dark matter. 

As a fact, the fermion generations in the standard model are identical, which transform the same, under the gauge symmetry and each generation is anomaly free. The number of fermion generations can in principle be arbitrary. All these might be a consequence of special weak-isospin group $SU(2)_L$ that its anomaly vanishes for every chiral fermion representation~\cite{anomaly}. By the new weak-isospin symmetry, the $SU(3)_L$ anomaly is nontrivial that is only cancelled if the number of generations is an integer multiple of three \cite{anomaly331}. Due to the contribution of exotic quarks along with ordinary quarks, QCD asymptotic freedom requires the number of generations lesser than or equal to five. So, the fermion generation number is three coinciding with the observations \cite{pdg}. 

Moreover, the fermion generations in the new model are non-universal that the third generation of quarks transforms under $SU(3)_L$ differently from the two others. This might provide a natural solution for the uncharacteristic heaviness of top quark \cite{longvan}. The quantization of electric charge is a consequence of fermion content under this new symmetry~\cite{ecq}. The model can by itself contain a Peccei-Quinn symmetry for solving the strong CP problem~\cite{palp}. The $B-L$ number behaves as a gauge charge (and $R$-parity results) since it does not commute and nonclosed algebraically with the 3-3-1 symmetry, which provides insights in the known 3-3-1 model \cite{dongdm,dongnew}. The neutrino masses, possible leptogenesis \cite{neutrino331} and dark matter \cite{dm331,dm331p,quiros,dongdm,inert331} have been extensively studied.            

As a result of the new $SU(3)_L\otimes U(1)_X$ gauge symmetry, the minimal interactions of the theory (including gauge interactions, minimal Yukawa Lagrangian and minimal scalar potential) put the relevant particles (known as wrong-lepton particles \cite{dongdm} or similar ones in other versions) in pairs, similarly to the case of superparticles in supersymmetry. Hence, the 3-3-1 model has been thought to give some candidates for dark matter \cite{dm331,dm331p,quiros}. However, the problem is how to suppress or evade the unwanted interactions (almost other than the minimal interactions) and the unwanted vacuums (come from neutral scalar candidates) that leads to the fast decay of dark matter (for detailed reviews, see \cite{dongdm,inert331}). 

It is easily realized that the new particles in the minimal 3-3-1 model \cite{331m} cannot be dark matter because they are either electrically charged or rapidly decayed, even for just minimal Lagrangian. The 3-3-1 model with right-handed neutrinos encounters the same issue \cite{inert331}. Even the lepton number symmetry was first regarded as a dark matter stability mechanism \cite{dm331p}, but it is quite wrong since the generation of neutrino masses violates the lepton number. To overcome the difficulty, Refs. \cite{quiros} introduced another lepton sector (the model was changed and called as the 3-3-1 model with left-handed neutrinos). In another approach \cite{dongdm}, a mechanism for dark matter stability based on $W$-parity, similarly to $R$-parity in supersymmetry, was given. However, this stability mechanism works only with the particle content of the 3-3-1 model with neutral fermions \cite{dongfla}. Hence, the issue of dark matter identification and its stability in the typical 3-3-1 models remain unsolved. 

If the $B-L$ charge is conserved, the typical 3-3-1 models are not self-consistent (since the $B-L$ and 3-3-1 symmetries are algebraically nonclosed as mentioned \cite{dongdm,dongnew}). This also applies for other continuous symmetries imposed such as $U(1)_G$ in \cite{quiros} that do not commute with the 3-3-1 symmetry. One way to keep the typical 3-3-1 models self-contained is that they have to possess explicitly-violating interactions of lepton number. [Notice that the lepton number is thus an approximate symmetry while the baryon number is always conserved and commuted with the 3-3-1 symmetry]. And, a theory for the dark matter in the typical 3-3-1 models must take this point into account. 

As a solution to the dark matter issue in the typical 3-3-1 models, we have proposed in the previous work \cite{inert331} that if one scalar triplet of the 3-3-1 model with right-handed neutrinos is inert ($Z_2$ odd) while all other fields are even, the remaining two scalar triplets (well-known as the normal scalar sector) will result an economical 3-3-1 model self-consistently \cite{ecn331}. This model provides appropriate masses for neutrinos besides the dark matter as resided in the inert triplet. In this work, we sift such outcome for the minimal 3-3-1 model. 

The minimal 3-3-1 model has traditionally been studied to be worked with three scalar triplets $\rho=(\rho^+_1,\ \rho^0_2,\ \rho^{++}_3)$, $\eta=(\eta^0_1,\ \eta^-_2,\ \eta^+_3)$, $\chi=(\chi^-_1,\ \chi^{--}_2,\ \chi^0_3)$ and/nor one scalar sextet $S=(S^0_{11},\ S^-_{12},\ S^+_{13},\ S^{--}_{22},\ S^0_{23},\ S^{++}_{33})$. The question is which scalars are inert, while the rest or a part of it---the normal scalar sector is appropriate for symmetry breaking, mass generation, and yielding a realistic model on both sides: mathematical and phenomenological. In this work, let us restrict our study for the cases with a minimal normal scalar sector so that the inert sector is enriched responsibly for dark matter. Looking in the literature, the reduced 3-3-1 model \cite{r331} seems to be a candidate. However, this model is encountered with a problem of large FCNCs which is experimentally unacceptable. As an alternative approach, we will indicate that the minimal 3-3-1 model can behave as a so-called ``simple 3-3-1 model'' that is based on only the two scalar triplets $\eta$ and $\chi$ (which is different from the reduced 3-3-1 model given in \cite{r331} due to the scalar and fermion contents). The model will be proved to be realistic rather than the previous version \cite{r331}.   

With the proposal of the simple 3-3-1 model, the rest of scalars ($\rho$, $S$), even the replications of $\eta,\chi$ as well as possible variants of all them including new forms, can be the inert sector ($Z_2$ odd) responsible for dark matter. However, the most basic cases that result for the desirable inert sector can be summarized as  
\ben
\item The triplet $\rho$ is inert ($S$ is suppressed). However, this candidate ($\rho^0_2$) cannot be a dark matter due to the direct dark matter detection constraints.
\item The sextet $S$ is inert ($\rho$ is suppressed). This sextet does not provide any realistic dark matter candidate similarly to the previous case. However, a variant of it with $U(1)_X$ charge, $X=1$, yields a triplet dark matter.    
\item Introduce an inert scalar triplet as the replication of $\eta$ ($\rho$ and $S$ are suppressed). We have a doublet dark matter. 
\item Introduce an inert scalar triplet as the replication of $\chi$ ($\rho$ and $S$ are suppressed). This case yields a singlet dark matter. 
\een Note that a combination of the cases above or the whole can be interplayed in a single theory based on the simple 3-3-1 model, but they will not be considered in the current work.     
     
The rest of this work is organized as follows. In Section \ref{model} we propose the simple 3-3-1 model. The identification of physical scalars, Goldstone and gauge bosons is given. The fermion masses, proton stability and FCNCs are also investigated. In Section \ref{implication}, the dark matter theories that are based on the simple 3-3-1 model are respectively presented. The dark matter candidates of the model with $\rho$ inert triplet and of $S$ inert sextet are analyzed to rule them out. We will also show that the models with inert triplets as replications of $\eta$ and $\chi$, respectively, and the model with $X=1$ inert scalar sextet can provide realistic candidates for dark matter. To be completed, in Section \ref{viabledm} we will give a particular evaluation of the important dark matter observables and compare them to the experimental data. Finally, we summarize our results and conclude this work in Section \ref{conclusion}.

\section{\label{model}Simple 3-3-1 model}   

We will re-examine the reduced 3-3-1 model~\cite{r331} and the minimal 3-3-1 model \cite{331m} that leads to a new and realistic 3-3-1 model with minimal lepton and scalar contents---the so-called simple 3-3-1 model. To make sure this point, the simple 3-3-1 model will be explicitly pointed out to be consistent with the data. By the new approach, the dark matter models are emerged to be studied in the next section.

\subsection{Proposal of the model}

The gauge symmetry of the considering model is given by $SU(3)_C\otimes SU(3)_L\otimes U(1)_X$, where the first factor is ordinary color group while the rest is the extension of the electroweak symmetry as mentioned. The fermion content which is anomaly free is defined as \cite{331m}   
\bea \psi_{aL} &\equiv & \left(\begin{array}{c}
               \nu_{aL}  \\ e_{aL} \\ (e_{aR})^c
\end{array}\right) \sim (1,3,0),\crn 
Q_{\al L}  &\equiv& \left(\begin{array}{c}
  d_{\al L}\\  -u_{\al L}\\  J_{\al L}
\end{array}\right)\sim (3,3^*,-1/3),\hs Q_{3L} \equiv \left(\begin{array}{c} u_{3L}\\  d_{3L}\\ J_{3L} \end{array}\right)\sim
 \left(3,3,2/3\right), \\ u_{a
R}&\sim&\left(3,1,2/3\right),\hs d_{a R} \sim \left(3,1,-1/3\right),\crn
J_{\al R} &\sim&
\left(3,1,-4/3\right),\hs J_{3R} \sim \left(3,1,5/3\right),\nn \eea where $a=1,2,3$
and $\al= 1,2$ are family indices. The quantum numbers in parentheses are given upon 3-3-1
symmetries, respectively. The electric charge operator takes the form $Q=T_3-\sqrt{3}T_8+X$, where $T_i\ (i=1,2,...,8)$ are $SU(3)_L$ charges, while $X$ is that of $U(1)_X$ (below, the $SU(3)_C$ charges will be denoted by $t_i$). The new quarks possess exotic electric charges as $Q(J_\al)=-4/3$ and $Q(J_3)=5/3$. 

Because the third generation of quarks as imposed transforms under $SU(3)_L$ differently from the first two generations, the FCNCs due to the new neutral gauge boson ($Z'$) exchange is more constrained that yields a low bound of some TeV for the 3-3-1 breaking scale or the $Z'$ mass \cite{longvan}. Such a new physics scale is possibly still in the well-defined region of the theory, limited below the Landau pole of around 5 TeV \cite{landau}. By contrast, if the first or second quark generation was arranged differently from the two others like the reduced 3-3-1 model \cite{r331}, the resulting theory would be ruled out by the large FCNCs, provided that the new physics enters below the Landau pole. Furthermore, the theory would be invalid (or inconsistent) if one tried to push the new physics scale far above the Landau pole in order to prevent the FCNCs \cite{longvan,longhuyen}. All these will also be studied in the last of this section.     

The model can work with only two scalar triplets \cite{r331}. Upon the proposed fermion content, let us impose, however, the following two scalar triplets \bea \eta = \left(\begin{array}{c}
\eta^0_1\\
\eta^-_2\\
\eta^{+}_3\end{array}\right)\sim (1,3,0),\hs \chi = \left(\begin{array}{c}
\chi^-_1\\
\chi^{--}_2\\
\chi^0_3\end{array}\right)\sim (1,3,-1),\label{vev2}
\eea with VEVs \bea \langle \eta\rangle = \fr{1}{\sqrt{2}}
\left(\begin{array}{c}
u\\ 0\\ 0\end{array}\right),\hs  \langle \chi\rangle =\fr{1}{\sqrt{2}}
\left(\begin{array}{c}
0\\ 0\\ w\end{array}\right).\eea 
This yields a dominant tree-level mass for top quark, while some lighter quarks that have no tree-level mass will get consistent masses via either effective interactions (shown below) or radiative corrections \cite{ecn331}. Otherwise, if the two scalar triplets like \cite{r331} which are $\chi$ and another triplet $\rho\sim (1,3,1)$ are retained for this model (in this case the $\eta$ is suppressed), it will result a vanishing tree-level mass for the top quark that is unnatural to be induced by such subleading quantum effect or effective theory.   

The original study in \cite{r331} gave a comment on the scalar triplets of this model, however the fermion content was never changed that would always face the large FCNC problems. In a recent research \cite{quiros1}, the fermion content was changed, but the scalar sector of the reduced 3-3-1 model was retained, which would be encountered with a vanishing top quark mass at the tree-level. Hence, those issues have naturally been solved by this proposal. In other words, all the ingredients as stated above recognize an unique 3-3-1 model distinguished from the previous versions such as the reduced and minimal 3-3-1 models \cite{r331,331m} due to the difference in the fermion and/or scalar contents. This is a new observation of this work, which is going to be called as the ``simple 3-3-1 model''.              

\subsection{Scalar sector}

The scalar potential of the model is given by 
\be V_{\mathrm{simple}}=\mu^2_1\eta^\dagger \eta + \mu^2_2 \chi^\dagger \chi + \la_1 (\eta^\dagger \eta)^2+\la_2 (\chi^\dagger \chi)^2+\la_3(\eta^\dagger\eta)(\chi^\dagger \chi) +\la_4(\eta^\dagger \chi)(\chi^\dagger\eta),\ee where $\mu_{1,2}$ have mass-dimensions while $\la_{1,2,3,4}$ are dimensionless. The VEVs $u,\ w$ are given from the potential minimization as 
\be u^2=\fr{2(2\la_2 \mu_1^2-\la_3\mu_2^2)}{\la^2_3-4\la_1\la_2},\hs w^2=\fr{2(2\la_1 \mu_2^2-\la_3\mu_1^2)}{\la^2_3-4\la_1\la_2}.\ee
To make sure that \ben \item The scalar potential is bounded from below (vacuum stability),  
\item The VEVs $u,\ \om$ are nonzero (for symmetry breaking and mass generation), 
\item The physical scalar masses are positive,\een 
the parameters satisfy   
\be \mu^2_{1,2}<0,\hs \la_{1,2,4}>0,\hs -2\sqrt{\la_1\la_2}<\la_3<\mathrm{Min}\left\{2\la_1\left({\mu_2}/{\mu_1}\right)^2,2\la_2\left({\mu_1}/{\mu_2}\right)^2\right\}.\ee In addition, the VEV $w$ breaks the 3-3-1 symmetry down to the standard model symmetry and provides the masses for new particles, while the VEV $u$ breaks the standard model symmetry as usual and gives the masses for ordinary particles. Therefore, to keep a consistency with the standard model, we impose $w\gg u$. 

Expanding $\eta,\ \chi$ around the VEVs, we get $\eta^T=(\fr{u}{\sqrt{2}}\ 0\ 0)+(\fr{S_1+i A_1}{\sqrt{2}}\ \eta^-_2\ \eta^+_3)$ and $\chi^T=(0\ 0\ \fr{w}{\sqrt{2}})+(\chi^-_1\ \chi^{--}_2\ \fr{S_3+i A_3}{\sqrt{2}})$. Hence, the physical scalar fields with respective masses are identified as follows  
\bea && h \equiv c_\xi S_1-s_\xi S_3,\hs m^2_h=\la_1 u^2+\la_2 w^2-\sqrt{(\la_1 u^2-\la_2 w^2)^2+\la^2_3 u^2 w^2}\simeq \fr{4\la_1\la_2-\la^2_3}{2\la_2}u^2,\crn
 && H \equiv s_\xi S_1 + c_\xi S_3,\hs m^2_{H}=\la_1 u^2+\la_2 w^2+\sqrt{(\la_1 u^2-\la_2 w^2)^2+\la^2_3 u^2 w^2}\simeq 2\la_2 w^2,\\
 && H^{\pm}\equiv c_{\theta}\eta^\pm_3+s_{\theta}\chi^\pm_1,\hs m^2_{H^\pm}=\fr{\la_4}{2}(u^2+w^2)\simeq \fr{\la_4}{2} w^2.\nn \eea
 Here, we have denoted $c_x=\cos(x),\ s_x=\sin(x),\ t_x=\tan(x)$, and so forth, for any $x$ angle. The $\xi$ is $S_1$-$S_3$ mixing angle, while the $\theta$ is that of $\chi_1$-$\eta_3$. They are obtained as  
 \be t_\theta=\fr{u}{w},\hs t_{2\xi}=\fr{\la_3 u w}{\la_2 w^2-\la_1 u^2}\simeq \fr{\la_3 u}{\la_2 w}.\ee 
 
The $h$ field is the standard model like Higgs boson, while $H$ and $H^\pm$ are new neutral and singly-charged Higgs bosons, respectively, which is unlike \cite{r331}. There are eight massless scalar fields $G_Z\equiv A_1$, $G_{Z'}\equiv A_3$, $G^\pm_{W}\equiv \eta_2^\pm$, $G^{\pm\pm}_{Y}\equiv \chi^{\pm\pm}_2$ and $G^\pm_X\equiv c_\theta \chi^\pm_1-s_\theta \eta_3^\pm$ that correspond to the Goldstone bosons of eight massive gauge bosons $Z$, $Z'$, $W^\pm$, $Y^{\pm\pm}$ and $X^\pm$ (see below). In the effective limit, $u\ll w$, we have 
 \bea \eta \simeq \left(\begin{array}{c}
\fr{u+h+i G_Z}{\sqrt{2}}\\
G^-_W \\
H^+\end{array}\right),\hs \chi \simeq \left(\begin{array}{c}
G^-_X\\
G^{--}_Y\\
\fr{w+H+i G_{Z'}}{\sqrt{2}}\end{array}\right).
\eea

\subsection{Gauge sector}

The covariance derivative is given by $D_\mu=\pa_\mu + i g_s t_i G_{i\mu} + i g T_i A_{i\mu} + i g_X X B_\mu$, where $g_s,\ g$ and $g_X$ are the gauge coupling constants, while $G_{i\mu},\ A_{i\mu}$ and $B_\mu$ are the gauge bosons, as associated with the 3-3-1 groups, respectively. 
On the other hand, in the next section we will introduce extra scalars that are odd under a $Z_2$ symmetry (so-called the ``inert'' scalars). However, the inert scalars do not give the masses for the gauge bosons because they have no VEV due to the $Z_2$ symmetry. Therefore, the gauge bosons of the model get masses from this part of the Lagrangian, $\sum_{\Phi = \eta,\chi}(D_\mu\langle \Phi \rangle)^\dagger (D^\mu\langle \Phi \rangle)$, which results as follows.   

The gluons $G_i$ are massless and physical fields by themselves. The physical charged gauge bosons with masses are respectively given by 
\bea &&W^{\pm}\equiv \fr{A_1\mp i A_2}{\sqrt{2}},\hs m^2_W=\fr{g^2}{4}u^2,\\
&&X^\mp \equiv \fr{A_4 \mp i A_5}{\sqrt{2}},\hs m^2_X=\fr{g^2}{4}(w^2+u^2),\\
&& Y^{\mp\mp}\equiv \fr{A_6 \mp i A_7}{\sqrt{2}},\hs m^2_Y=\fr{g^2}{4}w^2. \eea The $W$ is like the standard model $W$ boson that yields $u\simeq 246\ \mathrm{GeV}$. The new gauge bosons $X$ and $Y$ have large masses in $w$ scale, satisfying the relation $m^2_X=m^2_Y+m^2_W$ which contrasts to \cite{r331} and that in the economical 3-3-1 model \cite{ecn331}. 
     
The photon field $A_\mu$ as coupled to the electric charge operator is easily obtained, 
\be A_\mu = s_W A_{3\mu}+ c_W\left(-\sqrt{3}t_W A_{8\mu}+\sqrt{1-3t^2_W}B_\mu\right),\ee
where $s_W=e/g = t/\sqrt{1+4t^2}$, with $t=g_X/g$, is the sine of Weinberg angle \cite{dl}. The standard model $Z_\mu$ boson and the new neutral gauge boson $Z'_\mu$ can be given orthogonally to $A_\mu$ as follows \cite{dl}  
\bea
&&Z_\mu= c_W A_{3\mu}- s_W\left(-\sqrt{3}t_W A_{8\mu}+\sqrt{1-3t^2_W}B_\mu\right),\\
&& Z'_\mu=\sqrt{1-3t^2_W}A_{8\mu}+\sqrt{3}t_W B_\mu.
\eea    
The $A_\mu$ is a physical field ($m_A=0$) and decoupled, whereas there is a mixing between $Z$ and $Z'$ given by the squared-mass matrix of the form, 
\bea  \left(\begin{array}{cc}
m^2_Z & m^2_{ZZ'}\\
m^2_{ZZ'} & m^2_{Z'}
\end{array}\right), \eea where 
\be m^2_Z=\fr{g^2}{4c^2_W}u^2,\hs m^2_{ZZ'}=\fr{g^2\sqrt{1-4s^2_W}}{4\sqrt{3}c^2_W}u^2,\hs m^2_{Z'}=\fr{g^2[(1-4s^2_W)^2u^2+4c^4_W w^2]}{12 c^2_W(1-4s^2_W)}.\ee 
Therefore, we have two physical neutral gauge bosons (beside the photon), 
\be Z_1=c_\varphi Z -s_\varphi Z',\hs Z_2 = s_\varphi Z + c_\varphi Z', \ee with the mixing angle 
\be t_{2\varphi}= \fr{\sqrt{3}(1-4s^2_W)^{3/2}u^2}{2c^4_W w^2-(1+2s^2_W)(1-4s^2_W)u^2}\simeq \fr{\sqrt{3}(1-4s^2_W)^{3/2}}{2c^4_W}\fr{u^2}{w^2}.\ee  
 and their masses
 \bea m^2_{Z_1} &=& \fr{1}{2}[m^2_Z+m^2_{Z'}-\sqrt{(m^2_Z-m^2_{Z'})^2+4m^4_{ZZ'}}]\simeq \fr{g^2}{4c^2_W}u^2,\\
 m^2_{Z_2}&=& \fr{1}{2}[m^2_Z+m^2_{Z'}+\sqrt{(m^2_Z-m^2_{Z'})^2+4m^4_{ZZ'}}]\simeq \fr{g^2c^2_W}{3(1-4s^2_W)}w^2. \eea 
 Because of $\varphi \ll 1$, we have $Z_1\simeq Z$ and $Z_2\simeq Z'$. The $Z_1$ is the standard model like $Z$ boson, while $Z_2$ is a new neutral gauge boson with the mass in $w$ scale. The mixing between $Z$ and $Z'$ was not regarded in \cite{r331}. 
 
The contribution to the experimental $\rho$-parameter can be calculated as 
 \be \Delta \rho \equiv \fr{m^2_W}{c^2_W m^2_{Z_1}}-1\simeq \fr{m^4_{ZZ'}}{m^2_Z m^2_{Z'}}\simeq \left(\fr{1-4s^2_W}{2c^2_W}\right)^2\fr{u^2}{w^2}. \ee Taking $s^2_W=0.231$ and $\Delta \rho < 0.0007$ \cite{pdg}, we have $w>460$ GeV. Since the other constraints yield $w$ in some TeV, we conclude that the $\rho$-parameter is very close to one and in good agreement with the experimental data \cite{pdg}.                
        
\subsection{Fermion masses and proton stability}  
           
Again, the inert scalars as mentioned do not give the masses for fermions since they have no VEV and no renormalizable Yukawa interactions due to the $Z_2$ symmetry. Hence, the interactions that lead to the fermion masses are given only by the two scalar triplets above,  
\bea \mathcal{L}_Y&=&h^J_{33}\bar{Q}_{3L}\chi J_{3R}+ h^J_{\al \beta}\bar{Q}_{\al L} \chi^* J_{\beta R}\crn
&&+h^u_{3 a} \bar{Q}_{3L} \eta u_{aR}+ \fr{h^u_{\al a}}{\La} \bar{Q}_{\al L} \eta \chi u_{a R}\crn
&&+ h^d_{\al a} \bar{Q}_{\al L} \eta^* d_{a R} + \fr{h^d_{3 a}}{\La} \bar{Q}_{3L}\eta^*\chi^* d_{a R}  \crn
&&+h^e_{ab} \bar{\psi}^c_{aL} \psi_{bL}\eta + \fr{h'^e_{ab}}{\La^2}(\bar{\psi}^c_{aL}\eta\chi)(\psi_{bL}\chi^*)\crn 
&&+\fr{s^\nu_{ab}}{\La} (\bar{\psi}^c_{aL}\eta^*)(\psi_{bL} \eta^*)+H.c.,\eea where the $\La$ is a new scale (with the mass dimension) under which the effective interactions take place. It is easily checked that $h^e_{ab}$ is antisymmetric while $s^{\nu}_{ab}$ is symmetric in the flavor indices. The coupling $s^\nu$ explicitly violates the lepton number by two unit (as also needed for a realistic 3-3-1 model), while the other couplings $h$'s conserve this charge. Notice that the effective interactions for quark and neutrino masses start from five dimensions while for the charged leptons, it is from six dimensions.  

Let us remark on the properties of effective interactions. \ben \item {\it No evidence for a GUT and strength of effective interactions}: Since the perturbative property of the $U(1)_X$ interaction is broken as well as the Landau pole appears at a low scale of some TeV, the model has no origin from a more-fundamental theory such as GUTs at a higher energy scale. This contradicts to the case of the standard model and the 3-3-1 model with right-handed neutrinos. Therefore, we do not have such a GUT to compare and to say about the size of the effective interactions. \item {\it Smallness of neutrino masses}: The coupling $s^\nu$ violates lepton number, so it should be very small in comparison to the conserved ones $h$'s for charged leptons and quarks, $s^\nu\ll h$'s (since, by contrast, the conservation of lepton number implies $s^\nu=0$ but $h$'s~$\neq0$). Therefore, the five-dimensional interaction is reasonably to provide the small masses for neutrinos in spite of $\La \sim w$ in TeV order, which is unlike the canonical seesaw scale motivated by GUTs \cite{pdg} due to the above remark. [Notice that Ref. \cite{r331} discussed the cases with respect to five- or seven-dimensional interactions, despite the fact that all the effective interactions of this kind give comparable contributions with $\La\sim w$]. We conclude that the neutrino masses are generated to be naturally-small as a result of the mentioned approximate symmetry of lepton number, characterized by $\epsilon\equiv s^\nu/h\ll 1$ for all $h$'s. \item {\it Lepton parity and proton stability}: The lepton number of lepton triplet ($\psi$) components, for example, is $L=\mathrm{diag}(1,1,-1)$ which does not commute with the gauge symmetry. In fact, it is an approximate symmetry. Let us introduce a conserved symmetry as a remnant subgroup of the lepton number, \be P=(-1)^L,\ee so-called lepton parity. The lepton parity for the lepton triplet components is $P=\mathrm{diag}(-1,-1,-1)=-1$ and $P=\mathrm{diag}(1,1,1)=1$ for scalar triplets, quark triplets/antitriplets, $P=1$ for right-handed quark singlets, in spite of $L(J)=\pm 2$. Hence, the lepton parity always commutes with the gauge symmetry and conserved. It is just the mechanism for suppressing the effective interactions such as $\bar{\psi}^c_{1L}Q_{1L}\bar{u}^c_{1R}d_{1R}$ that lead to the proton decay, which is unlike the one in \cite{r331}. \een    

The mass Lagrangian of quarks and charged leptons takes the form $-\bar{f}_{aL}m^f_{ab}f_{bR}+H.c.$, where $f=J,\ u,\ d,\ e$. We have $m^J_{33}=-h^J_{33}w/\sqrt{2}$ as the mass of $J_3$, while $m^J_{\al\beta}=-h^J_{\al\beta} w/\sqrt{2}$ as the mass matrix of $J_{1,2}$. They all have large masses in $w$ scale. The mass matrices of $u$ and $d$ are respectively obtained as   
\bea m^u_{3a}=-h^u_{3a}\fr{u}{\sqrt{2}},\hs m^u_{\al a}=-h^u_{\al a} \fr{uw}{2\La},\hs m^d_{\al a}= -h^d_{\al a} \fr{u}{\sqrt{2}},\hs m^d_{3a}=h^d_{3a}\fr{uw}{2\La}.\eea Because of $\La \sim w$, the ordinary quarks $u$ and $d$ all get masses proportional to the weak scale $u=246$ GeV. For top quark, we have $m_t=-h^u_{33}\times 174$ GeV, provided that $h^u_{3a}$ is flavor-diagonal. Therefore, $m_t=173$ GeV if $h^u_{33}\approx 1$. On the other hand, the lighter quarks ($u,\ d,\ c,\ s,\ b$) can be explained by $h^{u}_{\al \beta}<1$, $h^d_{ab}<1$ as well as $w<\La$ which is more natural than the standard model. If the first or second generation of quarks was different under $SU(3)_L$, the mass of top quark would be $m_t=-h^u_{33}\fr{w}{\La}\times 123$ GeV, which is unnatural to achieve an experimental value of 173 GeV due to the fact that $h^u_{33}<1$ and $\fr{w}{\La}<1$ in the realm of perturbative theory. This issue is quite similar to the economical 3-3-1 model \cite{ecn331}. For the charged leptons, we derive
\be m^e_{ab}=\sqrt{2}u\left(h^e_{ab}+h'^e_{ba}\fr{w^2}{2\La^2}\right).\ee Since $\La\sim w$, the charged leptons have masses in the weak scale. Although $h^e$ is antisymmetric, $h'^e$ is a generic matrix in generation indices. Therefore, the charged lepton mass matrix takes the most general form that can provide consistent masses for the charged leptons in similarity to the case of the standard model. 

Finally, the mass Lagrangian of neutrinos is given by $-\fr{1}{2}\bar{\nu}^c_{aL} m^\nu_{ab}\nu_{bL}+H.c.$, where
\be m^\nu_{ab}= -s^\nu_{ab}\fr{u^2}{\La}.\ee To proceed further, let us give a comment of the neutrino masses of the model in \cite{r331} that look like $-\kappa'\fr{v^2_\rho}{2\La}\left(\fr{v_\chi}{\La}\right)^2$. This result that was given from a seven-dimensional interaction is similar in scale to ours as a fact that $v_\chi$ is close to $\La$. Rising in the dimension of effective interactions may not be a reason of smallness of the neutrino masses. Here, we have argued that the effective interaction responsible for the neutrino masses violates the lepton number as a character for the approximate symmetry of this charge (so that the 3-3-1 model is self-consistent). Whereas, all other mass operators do not have this property. On the other hand, our effective theory does not have a motivation from GUTs and for such case the effective interaction strengths such as $s^\nu$ are unknown. Hence, they just appear due to non-perturbative effects to reflect the observed phenomena. Indeed, using $\La = 5$ TeV, $u=246$ GeV and $m^\nu_{ab}\sim$ eV, we have $s^\nu_{ab}=\epsilon h \sim 10^{-10}$. Let us choose the Yukawa coupling of electron $h=h^e\sim 10^{-6}$. We get the lepton number violating parameter\be \epsilon\sim 10^{-4}.\ee The strength of the violating interaction for approximate lepton number is reasonably small in comparison to the ordinary interactions, and this may be the source why the neutrino masses are observed to be small.    

\subsection{FCNCs}

Let us give an evaluation of tree-level FCNCs that dominantly come from the gauge interactions. With the aid of $t=g_X/g$ and $X=Q-T_3+\sqrt{3}T_8$, the interaction of neutral gauge bosons is obtained by 
\be \mathcal{L}_{\mathrm{NC}}=-g\sum_\Psi \bar{\Psi}\ga^\mu[T_3A_{3\mu}+T_8A_{8\mu}+t(Q-T_3+\sqrt{3}T_8)B_\mu]\Psi, \ee 
where $\Psi$ runs over every fermion multiplet of the model. There is no FCNC coupled to $Q$ and $T_3$ since the flavors $\nu_{aL}$, $e_{aL}$, $e_{aR}$, $u_{aL}$, $u_{aR}$, $d_{aL}$, $d_{aR}$, $J_{\al L}$ and $J_{\al R}$ are respectively identical under these generators. Hence, the FCNCs happen only with $T_8$ that are given by 
\be \mathcal{L}_{T_8}=-g\sum_\Psi \bar{\Psi}\ga^\mu T_8(A_{8\mu}+t\sqrt{3}B_\mu)\Psi = -\fr{g}{\sqrt{1-3t^2_W}}\sum_{\Psi_L} \bar{\Psi}_L \ga^\mu T_8 \Psi_L Z'_\mu, \ee where we have used the identities $A_8+t\sqrt{3}B=(1/\sqrt{1-3t^2_W})Z'$ and $T_8(\Psi_R)=0$. In this case, there is no FCNC associated with the leptons and exotic quarks because the flavors $\nu_{aL}$, $e_{aL}$, $e_{aR}$ and $J_{\al L}$ correspondingly transform the same under $T_8$. Therefore, the FCNCs are only concerned to ordinary quarks ($u_{aL}$, $d_{aL}$) as a fact that under $T_8$ the third quark generation is different from the first two. The relevant part is 
\bea \mathcal{L}_{T_8} &\supset& -\fr{g}{\sqrt{1-3t^2_W}}[\bar{u}_{aL} \ga^\mu T_8(u_{aL}) u_{aL}+\bar{d}_{aL} \ga^\mu T_8(d_{aL}) d_{aL}] Z'_\mu\crn
&=& -\fr{g}{\sqrt{1-3t^2_W}}(\bar{u}_L\ga^\mu T_u u_L+\bar{d}_L\ga^\mu T_d d_L) Z'_\mu\crn
&=&-\fr{g}{\sqrt{1-3t^2_W}}[\bar{u}'_L\ga^\mu (V^\dagger_{uL}T_{u}V_{uL}) u'_L +\bar{d}'_L\ga^\mu (V^\dagger_{dL}T_{d}V_{dL}) d'_L] Z'_\mu, \eea
where $T_u=T_d=\fr{1}{2\sqrt{3}}\mathrm{diag(-1,-1,1)}$, $u=(u_1\ u_2\ u_3)^T$, $d=(d_1\ d_2\ d_3)^T$, $u'=(u\ c\ t)^T$ and $d'=(d\ s\ b)^T$. The $V_{uL}$ and $V_{dL}$ take part in diagonalizing the mass matrices of ordinary quarks, $u_{L}=V_{uL}u'_{L}$, $u_{R}=V_{uR}u'_{R}$, $d_{L}=V_{dL}d'_{L}$ and $d_{R}=V_{dR}d'_{R}$, so that $V^\dagger_{uL} m^u V_{uR}=\mathrm{diag}(m_u,m_c,m_t)$ and $V^\dagger_{dL} m^d V_{dR}=\mathrm{diag}(m_d,m_s,m_b)$. The CKM matrix is $V_{\mathrm{CKM}}=V_{uL}^\dagger V_{dL}$. 
Hence, the tree-level FCNCs are described by the Lagrangian,    
\be \mathcal{L}_{\mathrm{FCNC}}=-\fr{g}{\sqrt{1-3t^2_W}}(V^*_{qL})_{3i} \fr{1}{\sqrt{3}}(V_{qL})_{3j} \bar{q}'_{iL}\ga^\mu q'_{jL} Z'_\mu \hs (i\neq j),\ee where we have denoted $q$ as $u$ either $d$. 

With the above result, substituting $Z'=-s_\varphi Z_1+c_\varphi Z_2$, the effective Lagrangian for hadronic FCNCs can be derived via the $Z_{1,2}$ exchanges as 
\be \mathcal{L}^{\mathrm{eff}}_{\mathrm{FCNC}}=\fr{g^2[(V^*_{qL})_{3i} (V_{qL})_{3j}]^2}{3(1-3t^2_W)}\left(\fr{s_\varphi^2}{m^2_{Z_1}} + \fr{c_\varphi^2} {m^2_{Z_2}}\right) (\bar{q}'_{iL}\ga^\mu q'_{jL})^2.\ee The contribution of $Z_1$ is negligible since \be 
\fr{s^2_\varphi/m^2_{Z_1}}{c^2_\varphi/m^2_{Z_2}}\simeq \fr{(1-4s^2_W)^2}{4c^4_W}\fr{u^2}{w^2}\simeq 0.00244\times \fr{u^2}{w^2}\ll 1,\ee provided that  $s^2_W=0.231$ and $u\ll w$. Therefore, only $Z_2$ governs the FCNCs and we have
\be \mathcal{L}^{\mathrm{eff}}_{\mathrm{FCNC}}\simeq \fr{[(V^*_{qL})_{3i} (V_{qL})_{3j}]^2}{w^2} (\bar{q}'_{iL}\ga^\mu q'_{jL})^2.\label{zpcontri} \ee Interestingly enough, this interaction is independent of the Landau pole $1/(1-4s^2_W)$ (this is also an evidence pointing out that when the theory is encountered with the Landau pole, the effective interactions take place). It describes mixing systems such as $K^0-\bar{K}^0$, $D^0-\bar{D}^0$, $B^0-\bar{B}^0$ and $B^0_s-\bar{B}^0_s$, caused by pairs $(q'_i,q'_j)=(d,s),\ (u,c),\ (d,b),\ (s,b)$, respectively. The strongest constraint comes from the $K^0-\bar{K}^0$ system, given by \cite{pdg} 
\be \fr{[(V^*_{dL})_{31} (V_{dL})_{32}]^2}{w^2} < \fr{1}{(10^4\ \mathrm{TeV})^2}.\ee Assume that $u_a$ is flavor-diagonal. The CKM matrix is just $V_{dL}$ (i.e., $V_{\mathrm{CKM}}=V_{dL}$). Therefore, $|(V^*_{dL})_{31} (V_{dL})_{32}|\simeq 3.6\times 10^{-4}$ \cite{pdg} and we have 
\be w>3.6\ \mathrm{TeV}.\ee This limit is still in the perturbative region of the model \cite{landau} and is in good agreement with the recent bounds \cite{masszp}. 

By contrast, if the first or second generation of quarks is arranged differently from the two others under $SU(3)_L$, we have $|(V^*_{dL})_{11} (V_{dL})_{12}|\simeq |(V^*_{dL})_{21} (V_{dL})_{22}| \simeq 0.22$ \cite{pdg} for both the cases with the $K^0-\bar{K}^0$ system. Moreover, the new physics scale $w$ is bounded by the Landau pole, $w<5$ TeV, for example \cite{landau}. Hence, the effective coupling (\ref{zpcontri}) for the $K^0-\bar{K}^0$ system becomes $1.94\times 10^5/(10^4\ \mathrm{TeV})^2$ that is much greater than the above experimental bound by five order of magnitude. In other words, the experimental bound implies $w>2.2\times 10^3$ TeV, provided that the effective interaction (\ref{zpcontri}) works, which contradicts with the fact that the model in this region is invalid due to the limit of the Landau pole. Consequently, such cases should be ruled out due to the large FCNCs that are experimentally unacceptable. The third quark generation should be different from the first two.                     

\section{\label{implication} Implication for dark matter}

Let us note that the typical 3-3-1 models \cite{331m,331r} are generally supplied with three scalar triplets and/nor one scalar sextet. However, only the two scalar triplets among them (like the ones given above for the minimal 3-3-1 model or those in \cite{ecn331} for the 3-3-1 model with right-handed neutrinos) are sufficiently for symmetry breaking and mass generation. Hence, we would like to argue that the remaining scalar multiplets or similar ones (which have been discarded in the simple versions---the simple 3-3-1 model and the economical 3-3-1 model \cite{ecn331}) can behave as inert multiplets responsible for dark matter. The first work on this search was dedicated to the 3-3-1 model with right-handed neutrinos \cite{inert331}. 

For the case of minimal 3-3-1 model under consideration, the theoretical aspect and dark matter phenomenology will completely be distinguished from \cite{inert331} as well as the standard model extensions with a singlet, a doublet or a triplet scalar dark matter. For example, in the model of singlet dark matter, the dark matter interacts with the standard model matter via only the scalar portal. But, in this model, the singlet dark matter and the standard model matter can be coupled via the new gauge portal additionally. Also, the doublet and triplet dark matters can be communicated to the standard model matter by additional contributions of new scalars and new gauge bosons.             

\subsection{\label{inertrho} Simple 3-3-1 model with inert $\rho$ triplet}

We can introduce into the theory constructed above an extra scalar triplet as 
\be \rho = \left(
\begin{array}{c}
\rho^+_1 \\ \rho^0_2 \\ \rho^{++}_3
\end{array}\right)\sim (1,3,1). \ee This scalar triplet is a part of the minimal 3-3-1 model \cite{331m}. However, for the model under consideration we suppose that it transforms as an odd field under a $Z_2$ symmetry, $\rho\rightarrow -\rho$, whereas all other fields of the model are even. Therefore, the $\rho$ and its components (including the ones proposed below) are all called as inert fields/particles.    

The normal scalar sector $(\eta,\ \chi)$ which consists of the VEVs, the conditions for parameters and the physical scalars with their masses as obtained above remains unchanged \cite{inert331}. For the inert sector, $\rho$ has vanishing VEVs due to the $Z_2$ conservation. Moreover, the real and imaginary parts of electrically-neutral complex field $\rho^0_2=\fr{1}{\sqrt{2}}(H_\rho + i A_\rho)$ by themselves are physical fields. Any one of them can be stabilized if it is the lightest inert particle (LIP) among the inert particles resided in $\rho$ due to the $Z_2$ symmetry.  

Unfortunately, we can show that $H_\rho$ and $A_\rho$ cannot be a dark matter. Indeed, $H_\rho$ and $H_\rho$ are not separated (degenerate) in mass which leads to a scattering cross-section of $H_{\rho}$ and $A_\rho$ off nuclei due to the t-channel exchange by $Z$ boson. Such a large contribution has  already been ruled out by the direct dark matter detection experiments \cite{lhall}. 

This kind of model is not favored since it does not provide any dark matter. And, this is unlike the inert scalar triplet of the 3-3-1 model with right-handed neutrinos \cite{inert331}, even thought they play equivalently important roles for the typical 3-3-1 models \cite{331m,331r}.      

\subsection{\label{inerteta} Simple 3-3-1 model with $\eta$ replication}

An extra scalar triplet that is a replication of $\eta$ is defined as 
 \bea \eta' &=&  \left(\begin{array}{c}
\eta'^0_1\\
\eta'^-_2\\
\eta'^{+}_3\end{array}\right)\sim (1,3,0).\label{vev3}
\eea Here, the $\eta'$ and $\eta$ have the same gauge quantum numbers. However, they differ under a $Z_2$ symmetry. The $\eta'$ is assigned as an odd field under the $Z_2$, $\eta'\rightarrow - \eta'$, whereas the $\eta$ and all other fields of the simple 3-3-1 model are even.  

The scalar potential that is invariant under the gauge
symmetry and $Z_2$ is given by \bea
V&=&V_{\mathrm{simple}}+   \mu^2_{\eta'}
\eta'^\dagger \eta' +  x_1 (\eta'^\dagger \eta')^2+x_2(\eta^\dagger
\eta)(\eta'^\dagger \eta')+x_3(\chi^\dagger \chi)(\eta'^\dagger
\eta')\crn &&+x_4(\eta^\dagger \eta')(\eta'^\dagger \eta)
+x_5(\chi^\dagger \eta')(\eta'^\dagger\chi) +\fr 1 2
[x_{6}(\eta'^\dagger \eta)^2+H.c.]\label{the1} \eea
Here, $\mu_{\eta'}$ has the dimension of mass while $x_i$ $(i= 1,2,3,...,6)$ are dimensionless. All the parameters of the scalar potential are real, except that $x_{6}$ can be complex. But, the $x_{6}$'s phase can be eliminated by redefining the relative phases of
$\eta'$ and $\eta$. Therefore, this potential conserves the $CP$ symmetry. Moreover, the VEV of $\eta'$ vanishes due to the conservation of $Z_2$ symmetry. Hence, the $CP$ symmetry is also conserved by the vacuum. All the $x_6$, $u$ and $w$ can be considered to be real.   

Similarly to the previous case, the normal scalar sector ($\eta,\ \chi$) as identified above that includes the minimization conditions, the constraints on $u,\ w$, $\mu$'s, $\la$'s and the physical scalars with respective masses retains unchanged \cite{inert331}. To make sure that the scalar potential is bounded from below as well as the $Z_2$ symmetry is conserved by the vacuum, i.e. $\langle \eta'\rangle=0$, the remaining parameters of the potential satisfy \cite{inert331} \be 
\mu^2_{\eta'}>0,\hs x_{1,3}>0,\hs x_2+x_4\pm x_6>0.\ee Let us define $M^2_{\eta'}\equiv \mu^2_{\eta'}+ \fr{1}{2} x_2 u^2 + \fr{1}{2} x_3 w^2$ and $\eta'^0_1\equiv \fr{1}{\sqrt{2}}(H'_1+iA'_1)$. It is easily shown that the gauge states $H'_1$, $A'_1$, $\eta'^\pm_{2}$ and $\eta'^\pm_{3}$ by themselves are physical inert particles with the masses respectively given by  
\bea m^2_{H'_1} &=& M^2_{\eta'}+\fr 1 2 (x_4+x_6)u^2,\hs m^2_{A'_1}=M^2_{\eta'}+\fr 1 2 (x_4-x_6)u^2,\crn
m^2_{\eta'_2}&=&M^2_{\eta'},\hs m^2_{\eta'_3}=M^2_{\eta'}+ \fr 1 2 x _5 w^2. \eea The LIP responsible for dark matter is $H'_1$ if $x_6<\mathrm{Min}\{0,\ -x_4,\ (w/u)^2x_5-x_4\}$, or alternatively $A'_1$ if $x_6>\mathrm{Max}\{0,\ x_4,\ x_4- (w/u)^2x_5\}$. Let us consider the case $H'_1$ as the dark matter candidate (or a LIP). The $H'_1$ transforms as a doublet dark matter under the standard model symmetry which is similar to the case of the inert doublet model \cite{idmma}. However, the $H'_1$ has a natural mass in the $w$ scale of TeV range. Therefore, this model predicts the large mass region of a doublet dark matter \cite{lhall1}. Its relic density, direct and indirect detections can be calculated to fit the data \cite{dst}.

\subsection{\label{inertchi} Simple 3-3-1 model with $\chi$ replication} 
The $\chi$ replication has the form
\bea \chi' &=&  \left(\begin{array}{c}
\chi'^-_1\\
\chi'^{--}_2\\
\chi'^0_3\end{array}\right)\sim (1, 3,-1).\label{vev11}\eea 
Let us introduce a $Z_2$ symmetry so that $\chi'\rightarrow -\chi'$ while all other fields of the simple 3-3-1 model are even under this parity. 
The scalar potential that is invariant under the gauge symmetry and the $Z_2$ is given by \bea
V&=&V_{\mathrm{simple}}+ \mu^2_{\chi'} \chi'^\dagger \chi' + y_1 (\chi'^\dagger \chi')^2 +y_2(\eta^\dagger \eta)(\chi'^\dagger
\chi')+y_3(\chi^\dagger \chi)(\chi'^\dagger \chi')\crn
&&+y_4 (\eta^\dagger \chi')(\chi'^\dagger\eta)+y_5(\chi^\dagger
\chi')(\chi'^\dagger \chi) +\fr 1 2 [y_{6}(\chi'^\dagger
\chi)^2+H.c.] \eea 

Similarly to the previous model, we can take $y_{6}$, $u$ and $w$ as real parameters and the $CP$ symmetry is always conserved and unbroken by the vacuum. The normal scalar sector as obtained retains unchanged. The scalar potential is bounded from below and the $Z_2$ is conserved by the vacuum if we impose \be \mu^2_{\chi'}>0,\hs  y_{1,2}>0,\hs y_3+y_5\pm y_6>0.\ee  

By putting $M^2_{\chi'}\equiv \mu^2_{\chi'}+\fr 1 2 y_2 u^2+\fr 1 2 y_3 w^2$ and $\chi'^0_3\equiv \fr{1}{\sqrt{2}}(H'_3+iA'_3)$, we have the $H'_3$, $A'_3$, $\chi'^\pm_1$ and $\chi'^{\pm\pm}_2$ as physical inert scalar fields by themselves with corresponding masses,
\bea m^2_{H'_3}&=&M^2_{\chi'}+\fr 1 2 (y_5+y_6)w^2,\hs m^2_{A'_3}=M^2_{\chi'}+\fr 1 2 (y_5-y_6)w^2,\crn
m^2_{\chi'_2} &=& M^2_{\chi'},\hs m^2_{\chi'_1}=M^2_{\chi'}+\fr 1 2 y_4 u^2,  \eea which are all in the $w$ scale of TeV order.    

Depending on the parameter regime, $H'_3$ or $A'_3$ may be the LIP responsible for dark matter. Let us consider $H'_3$ as the LIP, i.e. $y_6<\mathrm{Min}\{0,\ -y_5,\ (u/w)^2y_4-y_5\}$. The $H'_3$ is a singlet dark matter under the standard model symmetry, similarly to the phantom of Silveira-Zee model \cite{zeemodel,singletdm}. However, its phenomenology is unique due to the interactions with the new gauge bosons and new Higgs bosons beside the standard model Higgs portal, which looks like the one in the 3-3-1 model with right-handed neutrinos \cite{inert331}. It has a natural mass in TeV range, and its relic density as well as the detection cross-sections can be calculated to compare with the data \cite{dst} (see also \cite{inert331} for the similar ones).      

\subsection{\label{sexdm} Simple 3-3-1 model with inert scalar sextet}
Since the inert scalar multiplets under consideration do not couple to fermions, their $U(1)_X$ charges are not fixed. However, these charges must be chosen so that at least one multiplet component is electrically-neutral for dark matter. Under this view, there are just three distinct inert scalar triplets $\rho$, $\eta'$ and $\chi'$ as already studied. However, there are only five inert scalar sextets since one of them contains up to two electrically-neutral components. In this work, we consider only the two sextets that are correspondingly embedded by the familiar scalar triplets with respective hyper-charges $Y=(+/-)1$ and $Y=0$ under the standard model symmetry: $(6,X)=(3,Y)\oplus(2,Y)\oplus(1,Y)$, where $Y=-\sqrt{3}T_8+X$ can be identified from the electric charge operator of the model.        

\subsubsection{Inert scalar sextet $X=0$}

Let us introduce the scalar sextet as often studied in the minimal 3-3-1 model \cite{331m} into the simple 3-3-1 model, 
\be S=\left(\begin{array}{ccc}
S^0_{11} & \fr{S^{-}_{12}}{\sqrt{2}} & \fr{S^+_{13}}{\sqrt{2}}\\ 
\fr{S^{-}_{12}}{\sqrt{2}} & S^{--}_{22} & \fr{S^0_{23}}{\sqrt{2}} \\
\fr{S^+_{13}}{\sqrt{2}} & \fr{S^0_{23}}{\sqrt{2}} & S^{++}_{33}
\end{array}
\right)\sim (1,6,0).\ee However, this sextet is odd under a $Z_2$ symmetry ($S\rightarrow -S$), while all other fields are even. Notice also that this sextet contains the scalar triplet with $Y=-1$ under the standard model symmetry similar to the one in the type II seesaw mechanism.   

The scalar potential is given by
\bea V &=& V_{\mathrm{simple}}+\mu^2_S \mathrm{Tr} S^\dagger S + z_1(\mathrm{Tr} S^\dagger S)^2+z_2\mathrm{Tr} (S^\dagger S)^2\crn
&& +(z_3\eta^\dagger \eta+z_4\chi^\dagger \chi)\mathrm{Tr} S^\dagger S + z_5\eta^\dagger S S^\dagger \eta+z_6\chi^\dagger S S^\dagger \chi\crn
&& + \fr 1 2 (z_7 \eta\eta S S +H.c.), \eea where the last terms can explicitly be written as $\eta\eta S S= \ep^{mnp}\ep^{qrs}\eta_m \eta_q S_{nr} S_{ps}$. To ensure that the potential is bounded from below as well as the $Z_2$ symmetry is conserved by the vacuum, i.e. $\langle S \rangle =0$, we impose 
\bea &&\mu^2_S>0,\hs z_1>0,\hs z_4>0, \hs z_1+z_2>0,\crn
&& z_3+z_5>0,\hs z_6+2z_4>0,\hs z_3\pm z_7>0.\eea Note that $z_7$ and the VEVs of $\eta$, $\chi$ can be chosen to be real due to the $CP$ conservation.    

Similarly to the above cases, the normal scalar sector as given remains unchanged. Let us put $M^2_S\equiv \mu^2_S+\fr 1 2 z_3 u^2 + \fr 1 2 z_4 w^2$, $S^0_{11}\equiv \fr{1}{\sqrt{2}}(H_S+iA_S)$ and $S^0_{23}\equiv \fr{1}{\sqrt{2}}(H'_S+iA'_S)$. The inert scalar sector yields the physical fields,
\bea && H_S,\hs A_S,\hs H'_S,\hs A'_S,\hs S^\pm_{12},\hs S^\pm_{13},\crn
&& H^{\pm\pm}_1=c_{\zeta} S^{\pm\pm}_{22}-s_{\zeta} S^{\pm\pm}_{33},\hs
H^{\pm\pm}_2=s_{\zeta} S^{\pm\pm}_{22}+c_{\zeta} S^{\pm\pm}_{33},\eea where $\zeta$ is the $S_{22}$-$S_{33}$ mixing angle defined by $t_{2\zeta}=\fr{2z_7}{z_6}\fr{u^2}{w^2}$. The masses of the inert particles are respectively given by
\bea m^2_{H_S}&=&m^2_{A_S}=M^2_S+\fr 1 2 z_5 u^2,\crn 
m^2_{H'_S} &=& M^2_S+\fr 1 4 z_6 w^2 -\fr 1 2 z_7 u^2,\hs m^2_{A'_S}=M^2_S+\fr 1 4 z_6 w^2 + \fr 1 2 z_7 u^2,\crn 
m^2_{S_{12}}&=&M^2_S+\fr 1 4 z_5 u^2, \hs m^2_{S_{13}}=M^2_S+\fr 1 4 z_5 u^2+\fr 1 4 z_6 w^2, \crn 
m^2_{H_{1,2}} &=& M^2_S+\fr 1 4 z_6 w^2 \mp \fr 1 4 \sqrt{z_6^2 w^4 + 4 z_7^2 u^4}. \eea 
All these masses are in the $w$ scale of TeV range. 

Depending on the parameter space, $H_S$, $A_S$, $H'_S$ and $A'_S$ may be dark matter candidates. However, $H_S$ and $A_S$ belong to the triplet under the standard model symmetry and they are degenerate in mass. Consequently, they have a t-channel exchange scattering off nuclei due to the contribution of $Z$ boson, which has already been ruled out by the direct dark matter detection experiments \cite{lhall}, similarly to those in the first dark matter model above. By contrast, $H'_{S}$ and $A'_S$ transform as doublets under the standard model symmetry and are separated in the masses. Unfortunately, they cannot be the LIP because both are much heavier than the $H_1$ field: $m^2_{H'_S(A'_S)}-m^2_{H_1}=\fr 1 4 \sqrt{z_6^2w^4+4z_7^2 u^4}-(+)\fr 1 2 z_7 u^2\simeq \fr 1 4 |z_6|w^2>0$. The $H'_S$ and $A'_S$ will rapidly decay that cannot be dark matter \cite{dst}. To conclude, the scalar sextet $S$ does not provide realistic dark matter candidates, which is similar to the case of the inert triplet model with corresponding scalar triplet as embedded in our sextet \cite{araki111}.

To resolve the mass degeneracy of the real and imaginary parts of neutral scalar field in the sextet (for the current model and even for the inert triplet model) as well as avoid the large direct dark matter detection cross-section, let us consider the following model.               

\subsubsection{Inert scalar sextet $X=1$}   
Let us introduce another sextet with $X=1$,  
\be \sigma=\left(\begin{array}{ccc}
\sigma^+_{11} & \fr{\sigma^{0}_{12}}{\sqrt{2}} & \fr{\sigma^{++}_{13}}{\sqrt{2}}\\ 
\fr{\sigma^{0}_{12}}{\sqrt{2}} & \sigma^{-}_{22} & \fr{\sigma^+_{23}}{\sqrt{2}} \\
\fr{\sigma^{++}_{13}}{\sqrt{2}} & \fr{\sigma^+_{23}}{\sqrt{2}} & \sigma^{+++}_{33}
\end{array}
\right)\sim (1,6,1).\ee This sextet is also odd under a $Z_2$ symmetry, whereas all the other fields are even. It is clear that the scalar triplet with $Y=0$ under the standard model symmetry has been embedded in the sextet. This scalar triplet has the gauge quantum numbers similarly to the standard model gauge triplet, and recently regarded for dark matter \cite{araki111} (see also \cite{masssplitting}).    

The scalar potential is given by
\bea V &=& V_{\mathrm{simple}}+\mu^2_\sigma \mathrm{Tr} \sigma^\dagger \sigma + t_1(\mathrm{Tr} \sigma^\dagger \sigma)^2+t_2\mathrm{Tr} (\sigma^\dagger \sigma)^2\crn
&& +(t_3\eta^\dagger \eta+t_4\chi^\dagger \chi)\mathrm{Tr} \sigma^\dagger \sigma + t_5\eta^\dagger \sigma \sigma^\dagger \eta+t_6\chi^\dagger \sigma \sigma^\dagger \chi\crn
&& + \fr 1 2 (t_7 \chi\chi \sigma \sigma +H.c.), \eea where all the couplings are real. The results of the normal scalar sector as obtained retain. The potential is bounded from below as well as the $Z_2$ symmetry is conserved by the vacuum if the new parameters satisfy  
\be \mu^2_\sigma>0,\hs 2t_1+t_2>0,\hs 2t_3+t_5>0,\hs t_4\pm t_7>0.\ee 

Denoting $M^2_\sigma\equiv \mu^2_\sigma+\fr 1 2 t_3 u^2 +\fr 1 2 t_4 w^2$ and $\sigma^0_{12}\equiv \fr{1}{\sqrt{2}}(H_\sigma+i A_\sigma)$, we have the physical fields, 
\bea && H_\sigma,\hs A_\sigma,\hs \sigma^\pm_{23},\hs \sigma^{\pm\pm}_{13},\hs \sigma^{\pm\pm\pm}_{33},\crn
&& H^{\pm}_{1}\equiv c_\delta \sigma^\pm_{11}-s_\delta \sigma^\pm_{22},\hs H^{\pm}_{2}\equiv s_\delta \sigma^\pm_{11}+c_\delta \sigma^\pm_{22},\eea
where $\delta$ is the mixing angle of $\sigma_{11}$-$\sigma_{22}$, defined by $t_{2\delta}=-\fr{2t_7}{t_5}\fr{w^2}{u^2}$. The corresponding masses for the fields are given by
\bea && m^2_{H_\sigma}=M^2_\sigma +\fr 1 4 t_5 u^2 -\fr 1 2 t_7 w^2,\hs m^2_{A_\sigma}=M^2_{\sigma} +\fr 1 4 t_5 u^2 +\fr 1 2 t_7 w^2,\crn
&& m^2_{\sigma_{23}}=M^2_\sigma +\fr 1 4 t_6 w^2,\hs m^2_{\sigma_{13}}=M^2_\sigma +\fr 1 4 t_5 u^2 +\fr 1 4 t_6 w^2,\hs m^2_{\sigma_{33}}=M^2_\sigma +\fr 1 2 t_6 w^2,\crn
&& m^2_{H_{1,2}}=M^2_\sigma +\fr 1 4 t_5 u^2 \mp \fr 1 4 \sqrt{t^2_5 u^4+4t^2_7 w^4}\simeq M^2_\sigma +\fr 1 4 t_5 u^2 \mp \fr 1 2 t_7 w^2 \mp\fr 1 8 \fr{t^2_5}{t_7}\fr{u^4}{w^2}, \eea which all have a natural size in the $w$ scale. 

It is noteworthy that the real and imaginary parts of the neutral scalar field of the standard symmetry triplet, $H_\sigma$ and $A_\sigma$, are separated in the masses as a result of the $\sigma$-$\chi$ interaction via the $t_7$ coupling. However, the masses of $H_\sigma$ and $H_1$ as well as those of $A_\sigma$ and $H_2$ are strongly degenerate, respectively, due to the $(u/w)^4\ll1$ suppression. As a fact, such small mass splittings are given by the tree-level contributions of the minimal scalar potential and are bounded by  
\bea |m_{H_1(H_2)}- m_{H_\sigma(A_\sigma)}| &\simeq& \left(\fr{t^2_5}{|t_7|}\right)\left(\fr{w}{m_{H_1(H_2)} + m_{H_\sigma(A_\sigma)}}\right)\left(\fr{3.6\ \mathrm{TeV}}{w}\right)^3 10\ \mathrm{MeV}\crn
 &\lesssim& 10\ \mathrm{MeV},\label{treedong1} \eea which is achieved due to $m_{H_1(H_2)} + m_{H_\sigma(A_\sigma)}\sim w$, $t_7\sim t_5 \sim 1$, 
$u\simeq 246\ \mathrm{GeV}$ and $w>3.6\ \mathrm{TeV}$. Further, the loop effects of the gauge bosons make the charged scalar masses larger than the neutral ones by an amount \cite{masssplitting}     
\be m_{H_1(H_2)}- m_{H_\sigma(A_\sigma)}\simeq 166\ \mathrm{MeV}.\label{loopstrumia1}\ee Combining the tree-level (\ref{treedong1}) and loop (\ref{loopstrumia1}) results, the charged scalars ($H_1,\ H_2$) are actually heavier than the neutral ones ($H_\sigma,\ A_\sigma$), respectively. [Note that the abnormal interactions such as $(\eta^\dagger T_i \eta) \mathrm{Tr}(\sigma^\dagger T_i \sigma)$ and $(\chi^\dagger T_i \chi) \mathrm{Tr}(\sigma^\dagger T_i \sigma)$ can also contribute to the mass differences of $H_\sigma(A_\sigma)$ and $H_1(H_2)$, respectively. But, these splitting effects are as small as the ones given by the minimal scalar potential, which can be neglected.] Therefore, either the $H_\sigma$ or the $A_\sigma$ can be regarded as the LIP responsible for dark matter. Without lost of generality, in the following let us consider $H_\sigma$ as the dark matter candidate, i.e. 
\be t_7>\mathrm{Max}\left\{0,\ -\fr 1 2 t_6,\ \fr 1 2 \left[t_5 (u/w)^2-t_6\right],\ \fr 1 2 \left[t_5 (u/w)^2-2 t_6\right]\right\}.\ee  

The notable consequences are that the contribution of $Z$ boson to the direct dark matter detection cross-section is suppressed because of the $H_\sigma$ and $A_\sigma$ mass splitting as well as the vanishing $H_\sigma A_\sigma Z$ interaction due to $T_3=Y=0$ for such scalar fields. The mass splitting of $H_\sigma$ and $A_\sigma$ is also necessary to prevent the $Z'$ contribution to such processes because the $Z'$ boson actually couples to $H_\sigma$ and $A_\sigma$, by contrast, due to $T_8\neq 0$ for the scalar fields. Indeed, if the contradiction happened ($t_7=0$), it would give rise to dangerous contributions naively proportional to $\sigma^{\mathrm{SI}}_{Z'}\sim \left(\fr{u}{w}\right)^4\sigma^{\mathrm{SI}}_{Z}\sim 10^{-43}\ \mathrm{cm}^2$ that is one up to two orders of magnitude larger than the best experimental bound $\sigma^{\mathrm{SI}}_{\mathrm{exp}}\sim 10^{-44}\ \mathrm{cm}^2-2\times 10^{-45}\ \mathrm{cm}^2$ \cite{xenon100}. Here, we have used $u=246\ \mathrm{GeV}$, $w=3.6-5\ \mathrm{TeV}$, and $\sigma^{\mathrm{SI}}_{Z}\sim 10^{-38}\ \mathrm{cm}^2$ as the cross-section for the case of the scalar triplet with $Y=-1$ and $Z$ exchange \cite{masssplitting}. 

\section{\label{viabledm} An evaluation of dark matter observables}                            

Along the above discussions, we have found the three dark matter candidates: a singlet scalar ($H'_3$), a doublet scalar ($H'_1$) and a triplet scalar ($H_\sigma$) under the standard model symmetry. And, they are absolutely stabilized due to the $Z_2$ symmetries as well as the fact that they are the LIPs. In fact, they could be viable dark matters because there always exist corresponding parameter regimes so that their relic densities, direct and indirect detection cross-sections are experimentally satisfied. Indeed, considering the parameter regimes that the candidates are lightest among the new particles of the corresponding models \cite{dongdm,inert331}, the dark matter observables are dominantly governed and set by the standard model particles, which have been well-established to be in agreement with the data \cite{singletdm,lhall1,araki111}. To be concrete, in the following we present for the case of the sextet dark matter.

Upon the aforementioned regime, the relic density for $H_\sigma$ includes only the processes that the candidate as well as the $H_1$ (co)annihilate into the standard model particles. They are governed by the Higgs and gauge portals with the corresponding interactions given by  
\bea V &\supset& \fr 1 4 (H^2_\sigma+2H^+_1 H^-_1) \left\{\left(t_3+\fr{t_5}{2}\right)h^2+\left[2t_3+t_5-\fr{\la_3}{\la_2}(t_4-t_7)\right] uh\right\},\\
\mathrm{Tr}[(D_\mu \sigma)^\dagger (D^\mu \sigma)] &\supset & g^2 H^2_\sigma W^+_\mu W^{-\mu}+ g^2 H_\sigma(H^+_1 W^-_\mu+H^-_1W^+_\mu)A^\mu_3 +\fr{g^2}{2}|H^+_1 W^-_\mu-H^-_1 W^+_\mu|^2 \crn
&&+ g^2H^+_1 H^-_1 A_{3\mu} A_3^\mu + igH^+_1\stackrel{\leftrightarrow}{\pa}_\mu H^-_1A^\mu_3 +[ig H_\sigma \stackrel{\leftrightarrow}{\pa}_\mu H^-_1 W^{+\mu} + H.c.],\eea where we have denoted $F_1\stackrel{\leftrightarrow}{\pa}_\mu F_2\equiv F_1(\pa_\mu F_2)-(\pa_\mu F_1)F_2$ for any $F_{1,2}$ fields, and $A_{3\mu}=s_W A_\mu + c_W Z_\mu$. The modification to the coupling of one $h$ with two inert particles is due to the $h$-$H$ mixing, which is at $u/w$ order. However, we have neglected the mixing effect of $Z$ with $Z'$ as well as the contribution of the new particles such as $H$ and $Z'$ because of $u^2\ll w^2$ and the above assumption for the dark matter candidate.      
 
\begin{figure}[!h]
\begin{center}
\includegraphics{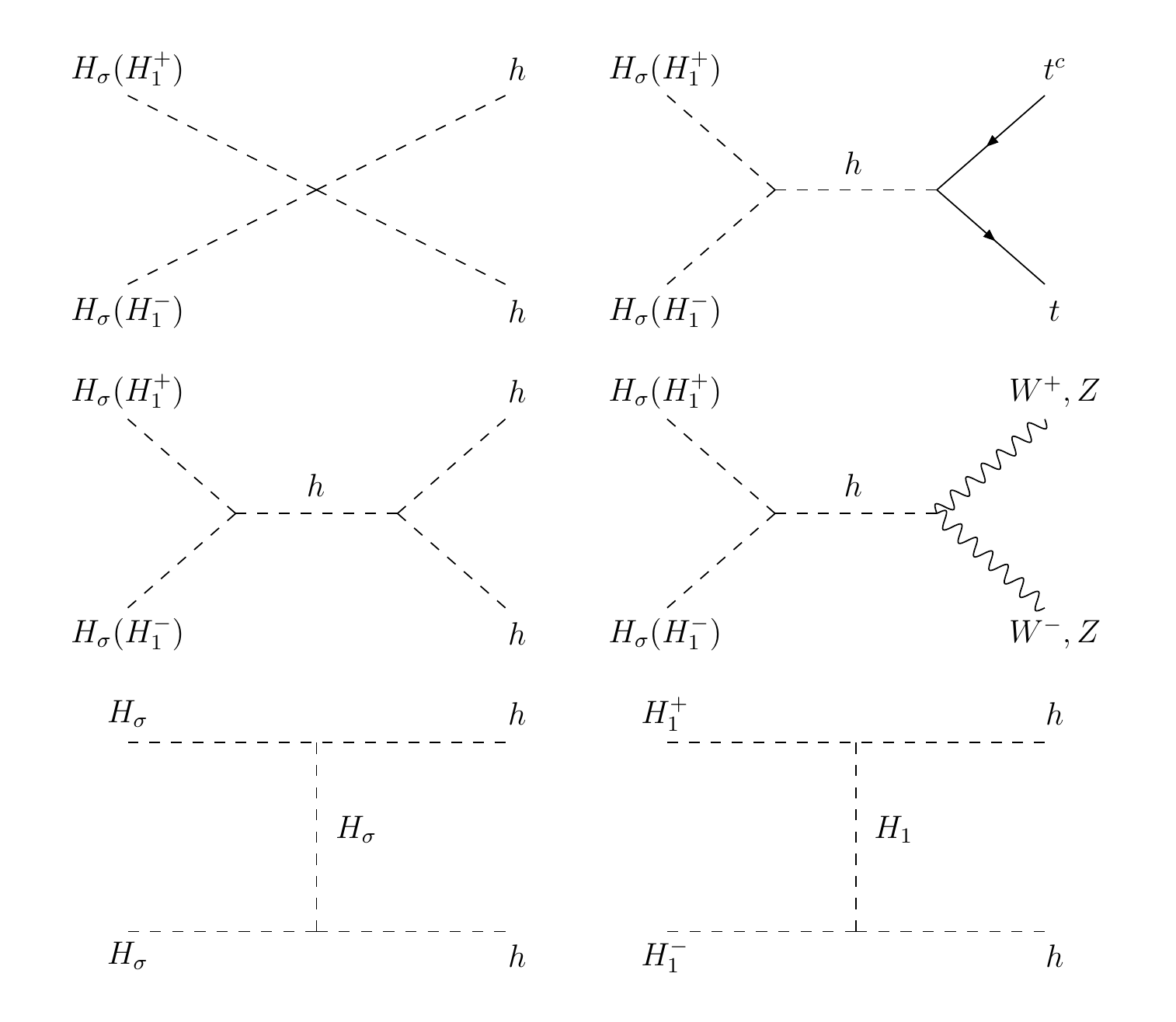}
\caption[]{\label{hportal} Contributions to $H_\sigma$ and/or $H^\pm_1$ annihilation via the Higgs portal when they are lighter than the new particles of the simple 3-3-1 model. There are additionally two $u$-channels that can be derived from the corresponding $t$-channels above.}
\end{center}
\end{figure}

\begin{figure}[!h]
\begin{center}
\includegraphics[width=14cm,height=18cm]{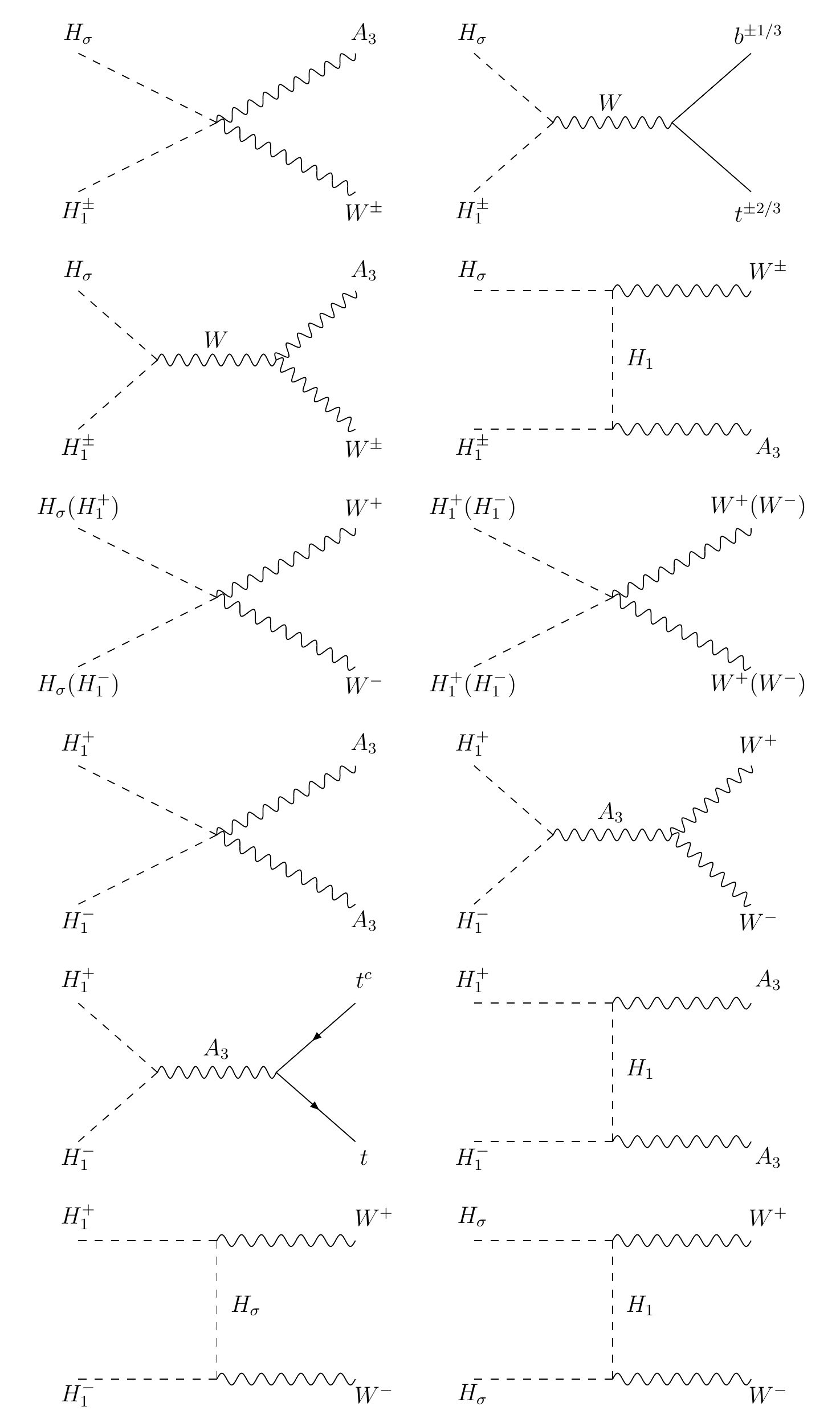}
\caption[]{\label{gportal} Contributions to $H_\sigma$ and/or $H^\pm_1$ annihilation via the gauge portal when they are lighter than the new particles of the simple 3-3-1 model. There remain the $u$-channel contributions for $H^+_1H^-_1\rightarrow A_3 A_3$ and $H_\sigma H_\sigma \rightarrow W^+W^-$, respectively, which can be extracted from the corresponding t-channel diagrams above.}
\end{center}
\end{figure}

There are various channels that might contribute to the relic density such as $H_\sigma H_\sigma\rightarrow hh,\ tt^c,\ W^+W^-,\ ZZ$ as well as the co-annihilations $H_\sigma H^\pm_1 \rightarrow ZW^\pm,\ A W^\pm,\ t^{\pm 2/3}b^{\pm  1/3}$ and $H^\pm_1 H^\mp_1\rightarrow hh,\ tt^c,\ W^+W^-,\ ZZ,\ ZA,\ AA$. They are given by the diagrams in Fig.~\ref{hportal} and Fig.~\ref{gportal} with respect to the Higgs and gauge portals, respectively. 
The annihilation cross-section times relative velocity is defined as $\sum_{ij}\sigma(H_i H_j\rightarrow \mathrm{SM\ particles})v_{ij}$, where $i,j=\sigma,1$ and $v_{ij}$ is the relative velocity of the two incoming particles $H_i$ and $H_j$. Using the limit $m_{H_{\sigma}}\simeq m_{H_1}\sim w\gg u\sim m_{\mathrm{SM}}$ (the relevant masses for the standard model particles) as well as the freeze-out temperature $T_F\simeq \fr{m_{H_{\sigma}}}{20}\ll m_{H_{\sigma}}$ as usual \cite{reliccal}, we obtain the leading order term for the effective, thermally-averaged annihilation cross-section times velocity,    
\be \langle \sigma v \rangle\simeq \fr{\al^2}{(150\ \mathrm{GeV})^2}\left[\left(\fr{2.3\ \mathrm{TeV}}{m_{H_{\sigma}}}\right)^2+\left(\fr{\la\times 0.782\ \mathrm{TeV}}{m_{H_\sigma}}\right)^2\right],\ee where the first term in the brackets comes from the gauge portal while the second one is due to the Higgs portal, $\la\equiv t_3+t_5/2$, in agreement with \cite{masssplitting}. For the above result, we have used $s^2_W=0.231$, $\al=1/128$. Note also that the quantity $\al^2/(150\ \mathrm{GeV})^2\simeq 1\ \mathrm{pb}$ has been factorized for a further convenience.    

The relic density can fit the data by this case if $\Omega h^2\simeq \fr{0.1\mathrm{pb}}{\langle \sigma v \rangle}\simeq 0.11$ (where the $h$ is the reduced Hubble constant) \cite{reliccal,pdg} that implies 
\be m_{H_\sigma}\simeq \sqrt{5.29+0.61\la^2}\ \mathrm{TeV}.\label{rdmm}\ee If the dark matter--scalar coupling is small $\la=t_3+t_5/2\ll1$, the gauge portal governs the annihilation processes of the dark matter. Simultaneously, the dark matter gets the right abundance if it has a mass $m_{H_\sigma}\simeq 2.3\ \mathrm{TeV}$. Otherwise, if the dark matter--scalar coupling is strength enough, $\la \gtrsim 1$, the Higgs portal gives equivalent contributions, even dominates over the gauge one. In this case, the dark matter mass depends on the $\la$ parameter as given above in order to recover the right abundance. Due to the limit by the Landau pole, say $m_{H_\sigma}<5\ \mathrm{TeV}$ (or equivalently $\la<5.68$ for the right abundance), the $H_\sigma$ can only contribute as a part of the total dark matter relic density, provided that the coupling $\la$ is large, $\la>5.68$. In other words, it is only a dark matter component coexisted with other potential candidates, which may be a singlet $H'_3$ and/or a doublet $H'_1$ as determined before.   

\begin{figure}[!h]
\begin{center}
\includegraphics{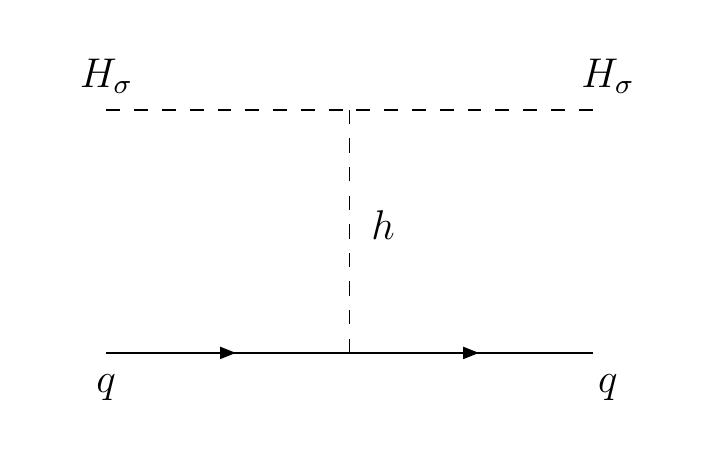}
\caption[]{\label{detection} Dominant contributions to $H_\sigma$-quark scattering.}
\end{center}
\end{figure}

The direct searches for the candidate $H_\sigma$ measure the recoil energy deposited by the $H_\sigma$ when it scatters off the nuclei of a large detector. This proceeds through the interaction of $H_\sigma$ with the partons confined in nucleons. Because the $H_\sigma$ is very non-relativistic, the process can be obtained by an effective Lagrangian as \cite{ddetect}
\be \mathcal{L_{\mathrm{eff}}}=2\la_q m_{H_\sigma} H_\sigma H_\sigma \bar{q}q,\ee where the scalar candidate has only spin-independent and even interactions (the interactions with gluons are loops induced that should be small). The above effective interaction is achieved by the $t$-channel diagram as mediated by the Higgs boson as Fig. \ref{detection}. 
It follows 
\be \la_q=\fr{\la' m_q}{2m_{H_\sigma}m^2_h},\hs \la'\equiv t_3+\fr{t_5}{2}-\fr{\la_3}{2\la_2}(t_4-t_7), \label{hsq}\ee where the scalar coupling $\la'$ that governs the scattering cross-section differs from the $\la$ that operates the annihilation cross-section. This separation is due to the term $\sim t_4-t_7$ raised as a result of the $h$-$H$ mixing. Hence, the relic density and the direct detection cross-section are obviously not correlated, which is a new observation of this work.            

The $H_\sigma$-nucleon scattering amplitude is obtained by summing over the quark level interactions multiplied by the corresponding nucleon form factors. Thus, the $H_\sigma$-nucleon cross-section takes the form,     
\be \sigma_{H_\sigma-N}=\fr{4m^2_{r}}{\pi}\la^2_N,\hs N= p,\ n, \ee where  
\bea m_r &\equiv& \fr{m_{H_\sigma}m_N}{m_{H_\sigma}+m_N}\simeq m_N,\hs \fr{\la_N}{m_N}=\sum_{u,d,s}f^N_{Tq}\fr{\la_q}{m_q}+\fr{2}{27}f^N_{TG}\sum_{c,b,t}\fr{\la_q}{m_q}\simeq 0.35\fr{\la'}{2m_{H_\sigma}m^2_h},\eea where $f^N_{TG}=1-\sum_{u,d,s}f^N_{Tq}$ as well as the $f^N_{Tq}$ values were given in \cite{john}. With $m_N=1$ GeV and $m_h=125$ GeV \cite{pdg}, we have    
\be \sigma_{H_\sigma-N} \simeq \left(\fr{2.494\la'\ \mathrm{TeV}}{m_{H_{\sigma}}}\right)^2 \times 10^{-44}\ \mathrm{cm}^2,\ee which coincides with the current experimental bound $\sigma_{H_\sigma-N}\simeq 10^{-44}\ \mathrm{cm}^2$, provided that $m_{H_\sigma}\simeq 2.494\la'\ \mathrm{TeV}$ in the TeV range \cite{pdg,xenon100}. Simultaneously, the $H_\sigma$ can get the right abundance by this case if we impose $\la'\simeq m_{H_\sigma}/(2.494\ \mathrm{TeV})\simeq \sqrt{0.85+0.098\la^2}\simeq 0.922\div 2$ with the help of (\ref{rdmm}) as well as $|\la|<5.68$ as mentioned. Of course, the direct detection cross-section can also be assigned to a smaller value if the coupling $\la'$ is appropriately chosen for each fixed dark matter mass.

\section{\label{conclusion}Conclusion}

Our aim was to look for a realistic 3-3-1 model with the minimal lepton and scalar contents in order to solve the dark matter problem of the minimal 3-3-1 model \cite{331m} under the guidance of the work in \cite{inert331}. However, there was not such a theory in the literature despite the fact that the reduced 3-3-1 model was introduced in \cite{r331}. And, for us it has been what to be investigated in this work.         

First of all, we have shown that even for a minimal 3-3-1 model with reduced scalar sector the third generation of quarks should transform under $SU(3)_L$ differently from the first two. This is due to the low limit of some TeV for the Landau pole. In addition, it is well-known that the mass corrections for some vanishing tree-level quark masses which come from quantum effects or effective interactions are subleading. Therefore, the reduced scalar sector must be $\eta$ and $\chi$ (no other case) so that the top quark appropriately gets a tree-level dominant mass. The simple 3-3-1 model that has been given by such minimal fermion and scalar contents is unique and entirely different from the previous one \cite{r331}.

We have also shown that there are eight Goldstone bosons correspondingly eaten by eight massive gauge bosons. There remain four physical Higgs bosons $h$, $H$ and $H^\pm$. Here the $h$ is like the standard model Higgs boson with mass in the weak scale while $H$ and $H^\pm$ are the new heavy Higgs bosons with masses in $w$ scale. Also, there is a small mixing between the standard model Higgs boson and the new one, $S_1$-$S_3$. Our model consists of only singly-changed Higgs bosons, not doubly-changed ones as in \cite{r331}.       

There are two new heavy charged gauge bosons with the masses in $w$ scale satisfying the relation $m^2_{X^{\pm}}=m^2_{Y^{\pm\pm}}+m^2_{W^\pm}$, which is unlike \cite{r331}. There is a mixing between the standard model $Z$ boson and the new neutral gauge boson $Z'$, which was neglected in \cite{r331}. The new physical neutral gauge boson $Z_2$ has a mass in $w$ scale. From the $W$ mass, we have $u\simeq 246$ GeV. On the other hand, from the constraint on the $\rho$ parameter, we get $w>460$ GeV. 

Because of the minimal scalar sector, some fermions have vanishing masses at the tree-level. However, they can get corrections coming from the effective interactions. The quarks get consistent masses via the five-dimensional effective interactions, while the charged leptons gain masses via four- and six-dimensional interactions. The  neutrino masses are generated to be naturally small as a consequence of approximate lepton number symmetry of the model. Notice that the model is only consistent by this way of the lepton charge. 

Although the lepton charge is an approximate symmetry, we can always find in the theory a conserved residual charge---the lepton parity $(-1)^L$. The conservation of lepton parity is just mechanism for the proton stability. Notice that the model always conserves the global baryon charge $U(1)_B$. This may also be regarded as a mechanism for the proton stability. 

We have calculated the hadronic FCNCs due to the exchange of $Z'$. It is interesting that the FCNCs are independent of the Landau pole. We have indicated that the strongest constraint coming from $K^0-\bar{K}^0$ system can be evaded provided that $w>3.6$ TeV. This value is still in the well-defined regime of the perturbative theory.  
        
The scalar multiplets other than the normal scalar sector of the simple 3-3-1 model, which include $\rho$ and $S$ as often studied in the minimal 3-3-1 model, $\eta'$ and $\chi'$ as the replications of the normal ones, the variants of $S$ such as $\sigma$ as well as the new forms, can be considered as the inert sectors providing dark matter candidates. We have shown that the simple 3-3-1 model with the inert scalar triplet $\rho$ does not contain any realistic dark matter. However, the simple 3-3-1 model with the $\eta$ or $\chi$ replication can yield a doublet dark matter $H'_1$ or a singlet dark matter $H'_3$, respectively. The simple 3-3-1 model with the inert scalar sextet $X=0$ does not provide any realistic dark matter. However, the model with the inert scalar sextet $X=1$ can give a triplet dark matter $H_\sigma$. The dark matter candidates as obtained can communicate with the standard model matter via the new Higgs and new gauge bosons besides the normal portals as in the ordinary inert triplet and inert doublet models as well as the Silveira--Zee model.                          

We have pointed out that the parameter spaces of the corresponding dark matter models can always contain appropriate parameter regimes so that the dark matter candidates as found are viable under the data. To be concrete, we have made an evaluation of the important dark matter observables for the sextet model that possesses the triplet scalar candidate ($H_\sigma$). This $H_\sigma$ gets a right abundance if it has a mass as $m_{H_\sigma}\simeq \sqrt{5.29+0.61\la^2}\ \mathrm{TeV}\simeq 2.3\div5\ \mathrm{TeV}$ for $|\la|<5.68$, where the annihilation cross-sections are operated by both the Higgs and gauge portals. The direct detection cross-section, which is governed by another scalar coupling $\la'$, is in good agreement with the experiments for the dark matter mass in TeV range. Taking the experimental bound as $\sigma_{H_\sigma-N}\simeq 10^{-44}\ \mathrm{cm}^2$, the dark matter mass is constrained to be $m_{H_\sigma}\simeq 2.494\la'$ TeV. The direct detection bound and right abundance are simultaneously satisfied if $\la'\simeq \sqrt{0.85+0.098\la^2}\simeq 0.922\div 2$ for $|\la|<5.68$.

\section*{Acknowledgments}

This research is funded by Vietnam National Foundation for Science and Technology Development (NAFOSTED) 
under grant number 103.01-2013.43.

\appendix


\begin{thebibliography}{99}

\bibitem{atlascms} G. Aad {\it et al.} (ATLAS Collaboration), Phys. Lett. B {\bf 716}, 1 (2012); S. Chatrchyan {\it et al.} (CMS Collaboration), Phys. Lett. B {\bf 716}, 30 (2012). 

\bibitem{pdg} J. Beringer {\it et al.} (Particle Data Group), Phys. Rev. D {\bf 86}, 010001 (2012).


\bibitem{331m} F. Pisano and V. Pleitez, Phys. Rev.  D {\bf 46}, 410 (1992);
P. H. Frampton, Phys. Rev. Lett. {\bf 69}, 2889 (1992); R. Foot,
O. F. Hernandez, F. Pisano and V. Pleitez, Phys. Rev. D {\bf 47},
4158 (1993).

\bibitem{331r} M. Singer, J. W. F. Valle and J. Schechter, Phys.
Rev. D {\bf 22}, 738 (1980); J. C. Montero, F. Pisano and V.
Pleitez, Phys. Rev. D {\bf 47}, 2918 (1993); R. Foot, H. N. Long
and Tuan A. Tran, Phys. Rev. D {\bf 50}, 34 (1994). 

\bibitem{ecn331} W. A. Ponce, Y. Giraldo
and L. A. Sanchez, Phys. Rev. D {\bf 67}, 075001 (2003); P. V.
Dong, H. N. Long, D. T. Nhung and D. V. Soa, Phys. Rev. D {\bf
73}, 035004 (2006); P. V. Dong and H. N. Long, Adv. High Energy
Phys. {\bf 2008}, 739492 (2008); P. V. Dong, Tr. T. Huong, D. T. Huong, and H. N. Long,
Phys. Rev. D {\bf 74}, 053003 (2006); P. V. Dong, H. N. Long, and D. V. Soa, Phys. Rev. D {\bf 73}, 075005 (2006); 
P. V. Dong, H. N. Long, and D. V. Soa, Phys. Rev. D {\bf 75}, 073006 (2007); P.V. Dong, D.T. Huong, M.C. Rodriguez, and H. N. Long, Nucl. Phys. B {\bf 772}, 150 (2007); P. V. Dong, H. T. Hung, and H. N. Long, Phys. Rev. D {\bf 86}, 033002 (2012).

\bibitem{r331} J. G. Ferreira Jr, P. R. D. Pinheiro, C. A. de S. Pires, and P. S. Rodrigues da Silva, Phys. Rev. D {\bf 84}, 095019 (2011). 

\bibitem{anomaly} S. Okubo, Phys. Rev. D {\bf 16}, 3528 (1977); J. Banks and H. Georgi, Phys. Rev. D {\bf 14}, 1158 (1976). 

\bibitem{anomaly331} See P. H. Frampton in \cite{331m}.

\bibitem{longvan} D. Ng, Phys. Rev. D {\bf 49}, 4805 (1994); D. G. Dumm, F. Pisano, and V. Pleitez, Mod. Phys. Lett. A {\bf 9}, 1609 (1994); H. N. Long and V. T. Van, J. Phys. G {\bf 25}, 2319 (1999).

\bibitem{ecq} F. Pisano, Mod. Phys. Lett A {\bf 11}, 2639 (1996);
A. Doff and F. Pisano, Mod. Phys. Lett. A {\bf 14},
1133 (1999); C. A. de S. Pires and O. P. Ravinez, Phys. Rev. D
{\bf 58}, 035008 (1998); C. A. de S. Pires, Phys. Rev. D {\bf 60},
075013 (1999); P. V. Dong and H. N. Long, Int. J. Mod. Phys. A
{\bf 21}, 6677 (2006).

\bibitem{palp} P. B. Pal, Phys. Rev. D {\bf 52}, 1659 (1995). 

\bibitem{dongdm} P. V. Dong, T. D. Tham, and H. T. Hung, Phys. Rev. D {\bf 87}, 115003 (2013).

\bibitem{dongnew} P. V. Dong, D. T. Huong, Farinaldo S. Queiroz, N. T. Thuy, arXiv:1405.2591 [hep-ph].

\bibitem{neutrino331} M. B. Tully and G. C. Joshi, Phys. Rev. D {\bf 64}, 011301 (2001); D. Chang and H. N. Long, Phys. Rev. D {\bf 73}, 053006 (2006); P. V. Dong and H. N. Long, Phys. Rev. D {\bf 77}, 057302 (2008); C. A. de S. Pires, F. Queiroz, and P. S. Rodrigues da Silva, Phys. Rev. D {\bf 82}, 065018 (2010); P. V. Dong, L. T. Hue, H. N. Long and D. V. Soa, Phys. Rev. D {\bf 81}, 053004 (2010); P. V. Dong, H. N. Long, D. V. Soa, and V. V. Vien, Eur.
Phys. J. C \textbf{71}, 1544 (2011); P. V. Dong, H. N. Long, C. H. Nam, and V. V. Vien, Phys. Rev. D {\bf 85}, 053001 (2012).

\bibitem{dm331} D. Fregolente and M. D. Tonasse, Phys. Lett. B {\bf 555}, 7 (2003); H. N. Long and N. Q. Lan, Europhys. Lett. {\bf 64}, 571 (2003); S. Filippi, W. A. Ponce, and L. A. Sanches, Europhys. Lett. {\bf 73}, 142 (2006).

\bibitem{dm331p} C. A. de S. Pires and P. S. Rodrigues da Silva, JCAP {\bf 0712}, 012 (2007). 

\bibitem{quiros} J. K. Mizukoshi, C. A. de S. Pires, F. S. Queiroz, and P. S. Rodrigues da Silva,  Phys. Rev. D {\bf 83}, 065024 (2011); J. D. Ruiz-Alvarez, C. A. de S. Pires, F. S. Queiroz, D. Restrepo, and P. S. Rodrigues da Silva, Phys. Rev. D {\bf 86}, 075011 (2012); S. Profumo and F. S. Queiroz, Eur. Phys. J. C {\bf 74}, 2960 (2014); C. Kelso, C. A. de S. Pires, S. Profumo, F. S. Queiroz, and P. S. Rodrigues da Silva, Eur. Phys. J. C {\bf 74}, 2797 (2014).  

\bibitem{inert331} P. V. Dong, T. Phong Nguyen, and D. V. Soa,  Phys. Rev. D 88, 095014 (2013). 

\bibitem{dongfla} P. V. Dong, L. T. Hue, H. N. Long
and D. V. Soa, Phys. Rev. D {\bf 81}, 053004 (2010); P. V. Dong, H. N. Long, D. V. Soa, and V. V. Vien, Eur.
Phys. J. C \textbf{71}, 1544 (2011); P. V. Dong, H. N. Long, C. H. Nam, and V. V. Vien, Phys. Rev. D {\bf 85}, 053001 (2012).

\bibitem{landau} Alex G. Dias, R. Martinez, and V. Pleitez, Eur. Phys. J. C {\bf 39}, 101 (2005).  

\bibitem{longhuyen} V. T. N. Huyen, T. T. Lam, H. N. Long, and V. Q. Phong, arXiv:1210.5833 [hep-ph].

\bibitem{quiros1} D. Cogollo, Farinaldo S. Queiroz, and P. Vasconcelos, 	arXiv:1312.0304 [hep-ph].

\bibitem{dl} P. V. Dong and H. N. Long, Eur. Phys. J. C {\bf 42}, 325 (2005).  

\bibitem{masszp} See, for examples, D. A. Gutierrez, W. A. Ponce, and L. A. Sanchez,  Eur. Phys. J. C {\bf 46}, 497 (2006); Y. A. Coutinho, V. S. Guimaraes, and A. A. Nepomuceno, Phys.Rev. D {\bf 87}, 115014 (2013). 

\bibitem{lhall} R. Barbieri, L. J. Hall, and V. S. Rychkov, Phys. Rev. D {\bf 74}, 015007 (2006).

\bibitem{idmma} N. G. Deshpande and E. Ma, Phys. Rev. D {\bf 18}, 2574 (1978).

\bibitem{lhall1} L. L. Honorez, E. Nezri, J. F. Oliver and M. H. G. Tytgat, JCAP {\bf 0702}, 028 (2007); M. Cirelli, N. Fornengo and A. Strumia, Nucl. Phys. B {\bf 753}, 178 (2006); T. Hambye, F. S. Ling, L. Lopez Honorez and J. Rocher, JHEP {\bf 07}, 090 (2009).  

\bibitem{dst} P. V. Dong, D. V. Soa and N. T. Thuy, ``The minimal 3-3-1 models for dark matter'', in preparation.

\bibitem{zeemodel} V. Silveira and A. Zee, Phys. Lett. B {\bf 161}, 136 (1985).  

\bibitem{singletdm} K.-M. Cheung, Y.-L. S. Tsai, P.-Y. Tseng, T.-C. Yuan, and A. Zee, JCAP {\bf 10}, 042 (2012); J. M. Cline, P. Scott, K. Kainulainen, and C. Weniger, Phys. Rev. D {\bf 88}, 055025 (2013). 

\bibitem{araki111} T. Araki, C. Q. Geng, and K. I. Nagao, Phys. Rev. D {\bf 83}, 075014 (2011). 

\bibitem{masssplitting} M. Cirelli {\it et al.} in \cite{lhall1}; M. Cirelli and A. Strumia, New J. Phys. {\bf 11}, 105005 (2009).

\bibitem{xenon100} E. Aprile {\it et al.} (XENON100 Collaboration), Phys. Rev. Lett. {\bf 109}, 181301 (2012). 

\bibitem{reliccal} G. Bertone, D. Hooper, and J. Silk, Phys. Rep. {\bf 405}, 279 (2005); J. Edsjo and P. Gondolo, Phys. Rev. D {\bf 56}, 1879 (1997); G. Jungman, M. Kamionkowski, and K. Griest, Phys. Rep. {\bf 267}, 195 (1996).

\bibitem{ddetect} G. Belanger, F. Boudjema, A. Pukhov, and A. Semenov, Comput. Phys. Commun. {\bf 180}, 747 (2009) [arXiv:0803.2360 [hep-ph]]. 

\bibitem{john} J. Ellis, A. Ferstl, and K. A. Olive, Phys. Lett. B {\bf 481}, 304 (2000). 

\end{thebibliography}
\end{document}